\documentclass[a4paper,11pt]{article}
\usepackage{pos}
\usepackage{METNet-defs}
\usepackage{subfig}
\usepackage{multirow}
\usepackage{footnote}
\usepackage{enumitem}
\makesavenoteenv{tabular}
\makesavenoteenv{table}
\graphicspath{{figures/}}

\title{Improving ATLAS Hadronic Object Performance with ML/AI Algorithms}
\ShortTitle{Improving ATLAS Hadronic Object Performance with ML/AI Algorithms}

\author[]{Benjamin Hodkinson, on behalf of the ATLAS Collaboration}
\renewcommand{\printHeadAuthors}{Benjamin Hodkinson}

\affiliation[]{University of Cambridge}

\emailAdd{ben.hodkinson@cern.ch}

\abstract{Hadronic object reconstruction is one of the most promising settings for cutting-edge machine learning and artificial intelligence algorithms at the LHC. In this contribution, selected highlights of ML/AI applications by ATLAS to particle and boosted-object identification, MET reconstruction and other tasks are presented.}

\FullConference{%
Presented at \textit{DIS2023: XXX International Workshop on Deep-Inelastic Scattering and Related Subjects}, Michigan State University, USA, 27-31 March 2023
}

\usepackage{lineno}

\begin{document}
% Remove when ready for final submission to conference:
%\linenumbers
%-----------------------------------------------------

\maketitle

\section{Introduction}
\label{sec:Intro}

Hadronic objects are ubiquitous in the proton--proton collision events recorded by the ATLAS detector \cite{PERF-2007-01} at the LHC. 
This includes hadronic jets, which are reconstructed from a large number of low-level calorimeter/track-based constituent objects, and missing transverse momentum (\etmiss), which involves every detector component and final-state object.
%which represents the total transverse momentum of undetected particles in an event, and its 
The complexity and abundance of jets and \ptmiss\ in the ATLAS dataset makes their reconstruction a promising setting for machine learning (ML) applications. 
This contribution introduces several recent developments in ATLAS which use ML to improve the performance of \ptmiss\ reconstruction \cite{METNET-PUBNOTE}, pion reconstruction \cite{ATL-PHYS-PUB-2022-040} and jet tagging \cite{ATL-PHYS-PUB-2022-039, ATL-PHYS-PUB-2021-029, Wtag_update}. 
These applications can broadly be separated into two categories:
\begin{enumerate}
	\item Regression of truth-level quantities from detector-level information.
	\item Classification of hadronic objects.
\end{enumerate}
\section{METNet: A combined \ptmiss\ working point}
\label{sec:metnet}

ATLAS employs several working points for \ptmiss\ reconstruction \cite{PERF-2016-07}, each of which is optimal for different event topologies and pile-up conditions. 
%predict \exmissterm{True}\ and \eymissterm{True}\ given the reconstructed \exyTmiss\ for four working points, plus additional information characterising pile-up and event topology. 
%The working points considered here are: \textit{Loose}, \textit{Tight}, \textit{Tighter} and \textit{Tenacious}. These have increasingly strict selections on hadronic jets which increases pile-up rejection but decreases hard-scatter jet acceptance.
\metnet\ \cite{METNET-PUBNOTE} is a neural network (NN) designed to pick and combine the reconstructed \ptmiss\ for each working point into a single \ptmiss\ estimate. 
This is achieved by regressing particle-level (`true') \ptmiss\ given the detector-level \ptmiss\ predictions for each working point and information characterising pile-up and event topology. %, such as number of primary vertices and mean number of interactions per bunch crossing. 
The NN is trained on a mixture of \ttbar\ and di-boson Monte Carlo (MC) events. 
The performance of two iterations of \metnet\ is presented here: one trained using the Huber loss \cite{Huber} function, and another including an additional Sinkhorn \cite{cuturi2013sinkhorn} contribution to the loss to reduce an observed negative bias, denoted METNet (Sk). 

\begin{figure}[t]
	\centering
	\subfloat[][]{
		\includegraphics[width=0.45\textwidth]{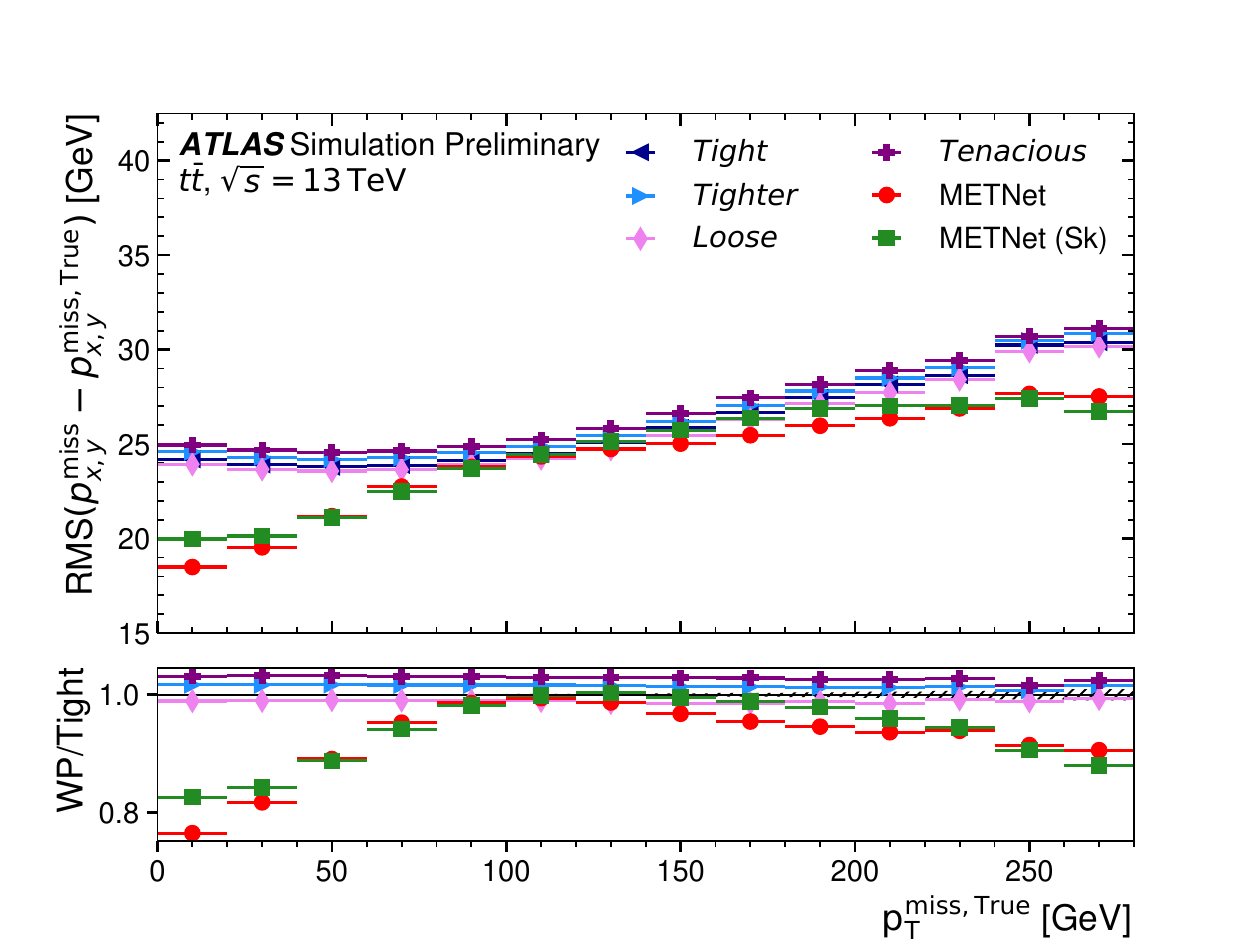}
		\label{fig:ttbar_TruthMETRMS}}
	\qquad
	\subfloat[][]{
		\includegraphics[width=0.45\textwidth]{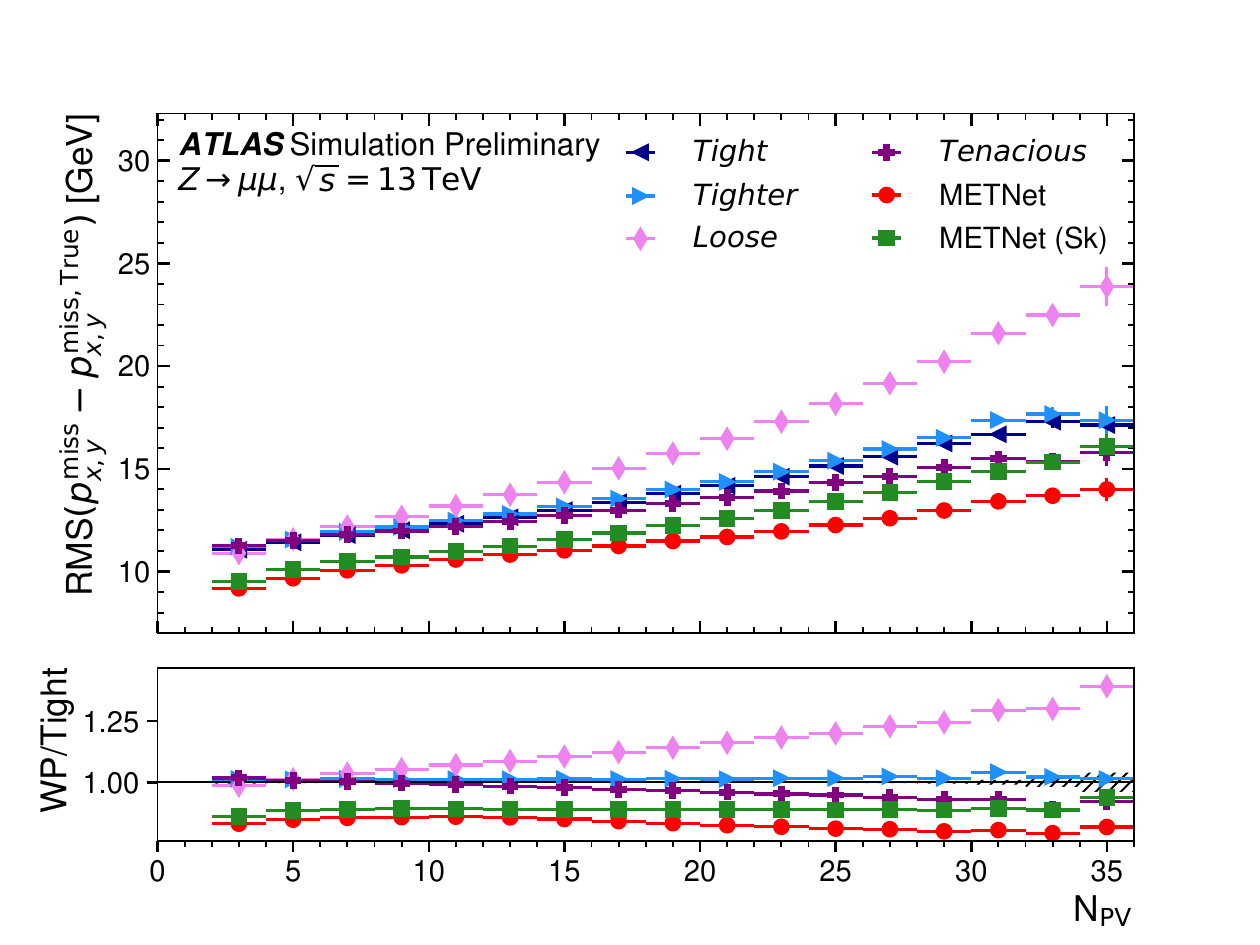}
		\label{fig:Zmm_NPVRMS}}
	\caption{Root-mean-square of the difference between the predicted value of \exymiss\ and \exymissterm{True} for \metnet\ and the  current \etmiss\ working points, for (a) \ttbar\ events in bins of \etmissterm{True}, (b) \Zmm\ events in bins of number of primary vertices. Plots reproduced from Reference~\cite{METNET-PUBNOTE}.}
	\label{fig:metnet}
\end{figure}

%Figure~\ref{fig:ttbar} shows the performance of \metnet\ for \ttbar\ events. 
%METNet has improved resolution compared to the \etmiss\ working points, indicated by the smaller root-mean-square error (RMSE) in Figure~\ref{fig:ttbar_TruthMETRMS}. 
%A negative bias is seen in Figure~\ref{fig:ttbar_response} (which shows the response \etmiss/\etmissterm{True}). This bias is reduced for METNet (Sk). 
%Figure~\ref{fig:RMS} shows the RMSE for \subref{fig:WW_TruthMETRMS} \WW\ and \subref{fig:Zmm_NPVRMS} \Zmm\ events, in bins of \etmissterm{True} and number of primary vertices respectively. 
%METNet has improved RMSE compared to current working points, showing an ability to generalise to topologies which were not seen during training.

Figure~\ref{fig:metnet} shows the root-mean-squared error (a metric for resolution) of \metnet\ and several current working points for \subref{fig:ttbar_TruthMETRMS} \ttbar\ and \subref{fig:Zmm_NPVRMS} \Zmm\ events, in bins of \etmissterm{True} and number of primary vertices respectively. 
METNet has improved resolution for both topologies and shows an ability to generalise to topologies such as \Zmm\ which were not seen during training.

%\begin{figure}[h]
%	\centering
%	\subfloat[][\ttbar\ resolution]{
%		\includegraphics[width=0.45\textwidth]{METNet_results/ttbar/Truth_MET_RMSxy.pdf}
%		\label{fig:ttbar_TruthMETRMS}}
%	\qquad
%	\subfloat[][\ttbar\ response]{
%		\includegraphics[width=0.45\textwidth]{METNet_results/ttbar/response.pdf}
%		\label{fig:ttbar_response}}
%	\caption{Comparison of \metnet\ and the  current \etmiss\ working points for \ttbar\ events: (a) Root-mean-square of the difference between the predicted value of \exymiss\ and \exymissterm{True} (a measure of resolution) in bins of \etmissterm{True}, (b) Response of the predicted \etmiss. Plots reproduced from Reference~\cite{METNET-PUBNOTE}.}
%	\label{fig:ttbar}
%\end{figure}

%\begin{figure}[h]
%	\centering
%	\subfloat[][\WW resolution]{
%		\includegraphics[width=0.45\textwidth]{METNet_results/WW/Truth_MET_RMSxy.pdf}
%		\label{fig:WW_TruthMETRMS}}
%	\qquad
%	\subfloat[][\Zmm\ resolution]{
%		\includegraphics[width=0.45\textwidth]{METNet_results/Zmm/NPV_RMSxy.pdf}
%		\label{fig:Zmm_NPVRMS}}
%	\caption{Root-mean-square of the difference between the predicted value of \exymiss\ and \exymissterm{True} for (a) \WW\ events in bins of \etmissterm{True}, (b) \Zmm\ events in bins of number of primary vertices. Plots reproduced from Reference~\cite{METNET-PUBNOTE}.}
%	\label{fig:RMS}
%\end{figure}

\begin{figure}[b]
	\centering
	\subfloat[][]{
		\includegraphics[width=0.5\textwidth]{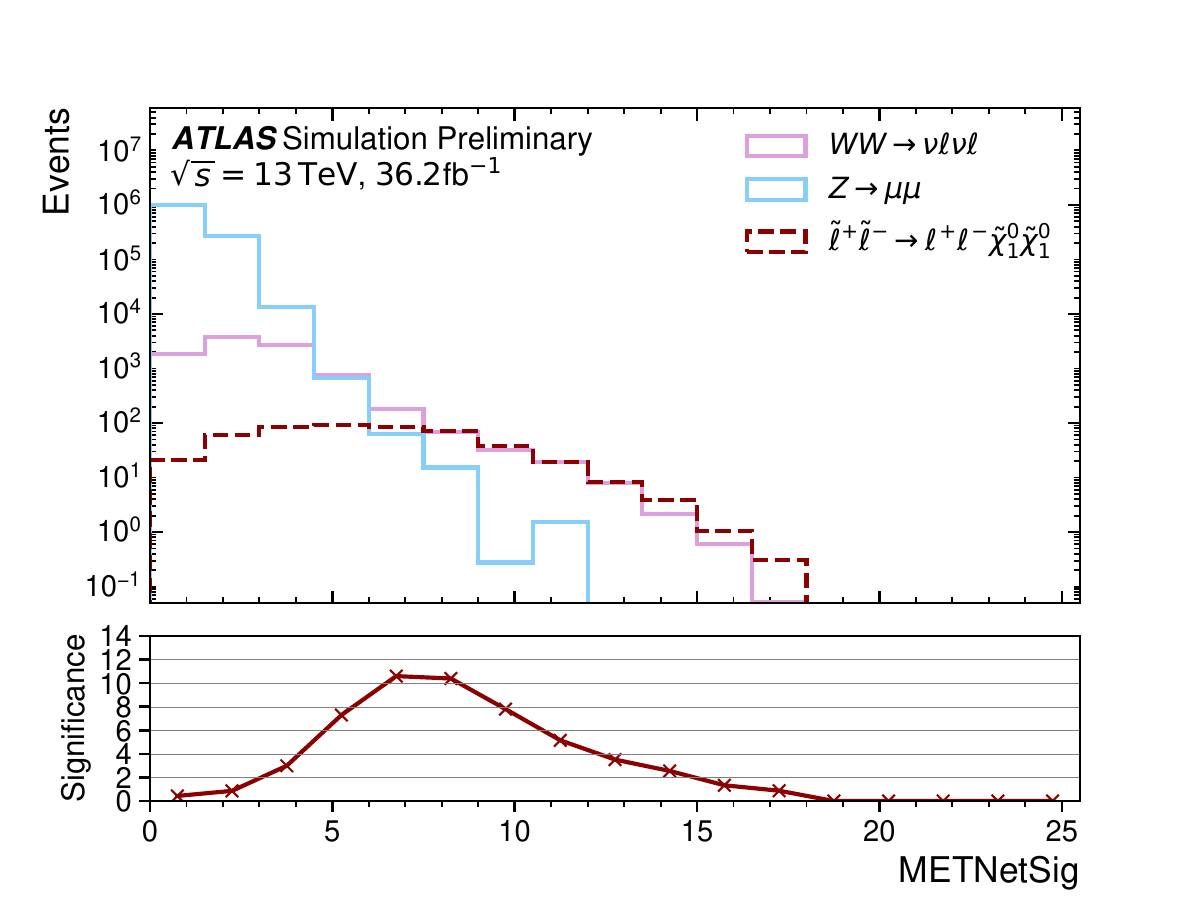}
		\label{fig:METNetSig}}
	\subfloat[][]{
		\includegraphics[width=0.5\textwidth]{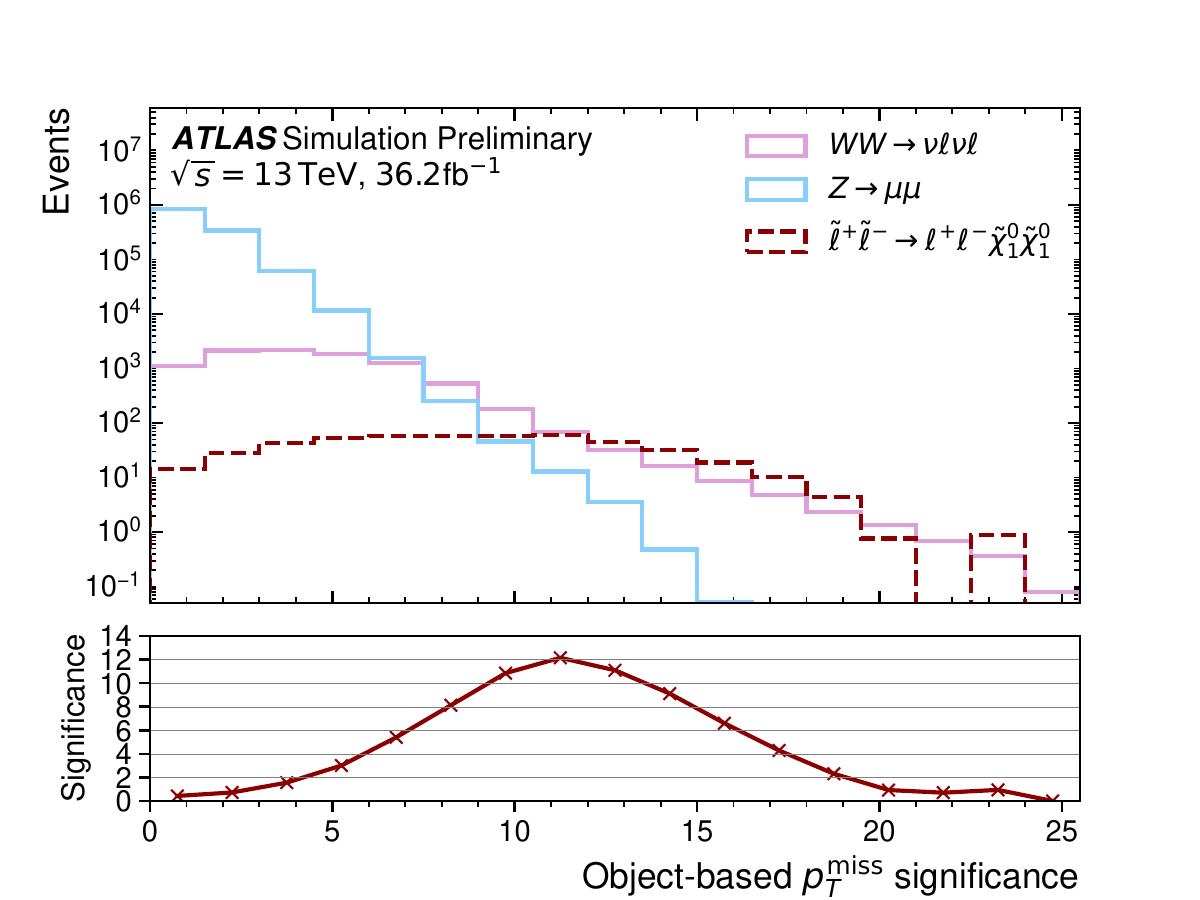}
		\label{fig:ObjMETSig}}
	\caption{Distributions of (a) \metnetsig\ and (b) object-based \etmiss\ significance for a supersymmetric signal sample and \(WW\) and \Zmm\ backgrounds. 
		%Events are weighted to 36.2 fb$^{-1}$, corresponding to the 2015+2016 ATLAS proton--proton collision dataset luminosity. 
		The lower panel shows the signal significance for a lower-bound selection at each $x$-axis bin value. Plots reproduced from \cite{METNET-PUBNOTE}.}
	\label{fig:METsig}
\end{figure}

The variable \etmiss\ significance \cite{ATLAS-CONF-2018-038} is also used in ATLAS to separate processes with `real' \ptmiss\ (from genuine invisible particles, such as neutrinos) and `fake' \ptmiss\ (from detector mis-measurement).
\metnet\ is extended to produce a `confidence' $\sigma$ as well as a central \ptmiss\ prediction by using the Gaussian negative log-likelihood (GNLL) loss. %\cite{Nix1994}
%\begin{equation}
%\(
%\mathcal{L}_{\text{GNLL}} = \log\sigma + 0.5\left(\frac{y-\hat{y}}{\sigma}\right)^2
%\mathcal{L}_{\text{GNLL}} = \log\sigma_{x,y} + 0.5\left(\frac{p_{x,y}^{\text{miss, NN}}- p_{x,y}^{\text{miss, True}}}{\sigma_{x,y}}\right)^2
%\).
%\end{equation}
%where \(y\) is \(p_{x,y}^{\text{miss, True}}\) and \(\hat{y}\) the NN's predicted value as before, and \(\sigma\) is an additional output variable which represents the network's confidence. 
A machine learning-based \ptmiss\ significance variable is then defined as METNetSig\( = p_{\text{T}}^\text{miss, NN} / \sigma\).
Figure~\ref{fig:METsig} shows \subref{fig:METNetSig} \metnetsig\ and \subref{fig:ObjMETSig} object-based \etmiss\ significance \cite{ATLAS-CONF-2018-038} (the current ATLAS state-of-the-art) for a supersymmetric signal process plus two Standard Model backgrounds.  
\metnetsig\ shows the ability to separate real and fake \ptmiss\ and has similar behaviour to object-based \etmiss\ significance.

%This contribution presents the performance of METNet and METNetSig - variables defined using the outputs of a neural network - compared to current ATLAS methods.
%These variables are calculated from outputs of a neural network trained on reconstructed \exymiss-terms for four established \ptmiss\ reconstruction working points, plus additional information to characterise event topology and levels of pile-up.
%METNet has significantly improved resolution for a range of topologies, including those not seen during training. Including a Sinkhorn contribution to the loss function reduces an observed negative bias in the NN's predictions. METNetSig shows similar behaviour per-topology to object-based \ptmiss significance, and can distinguish between real \ptmiss\ and fake \ptmiss. 

\section{Pion reconstruction}
\label{sec:pions}

\begin{figure}[h]
	\centering
	\subfloat[][]{
		\includegraphics[width=0.45\textwidth]{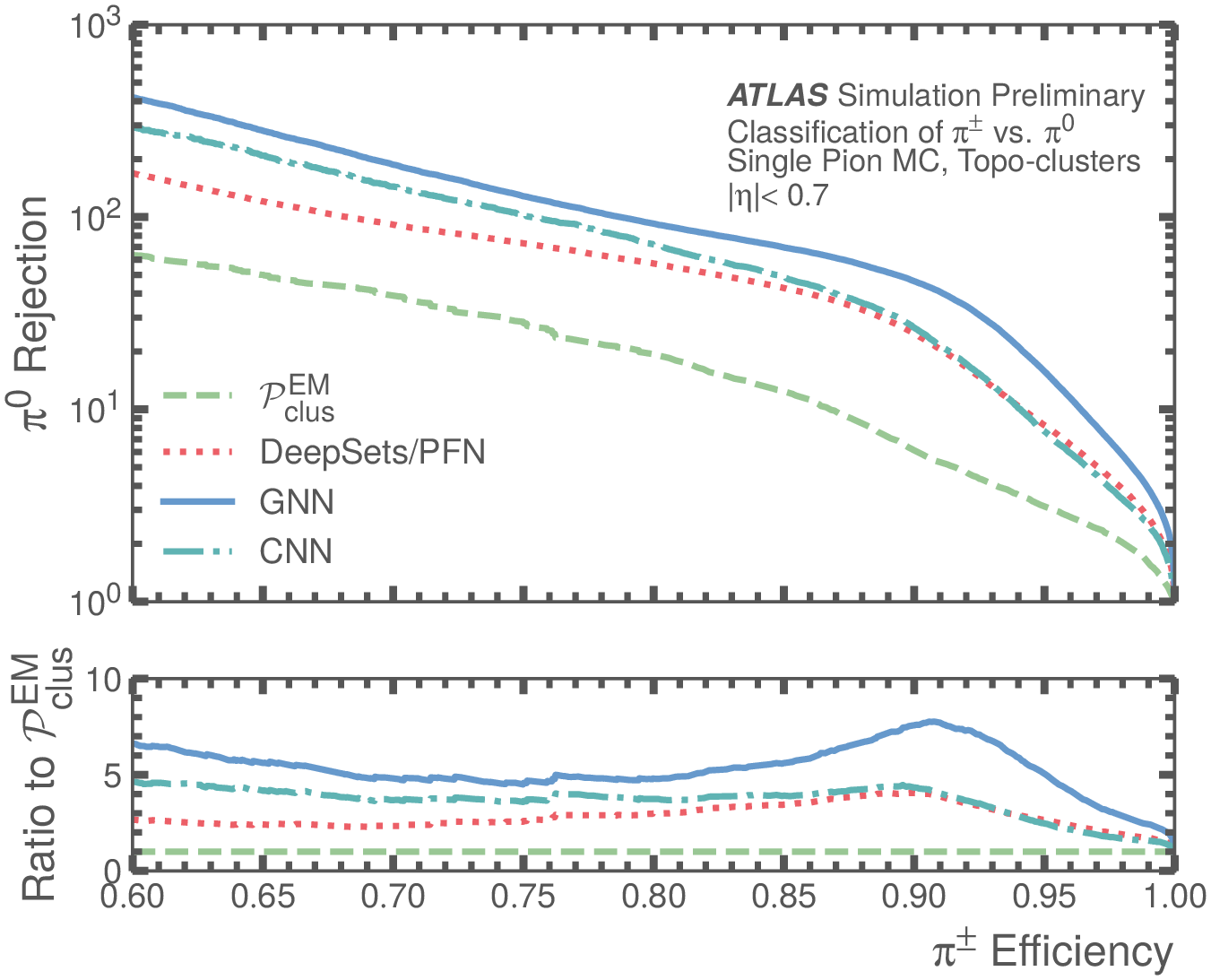}
		\label{fig:pion_class}}
	\qquad
	\subfloat[][]{
	\includegraphics[width=0.45\textwidth]{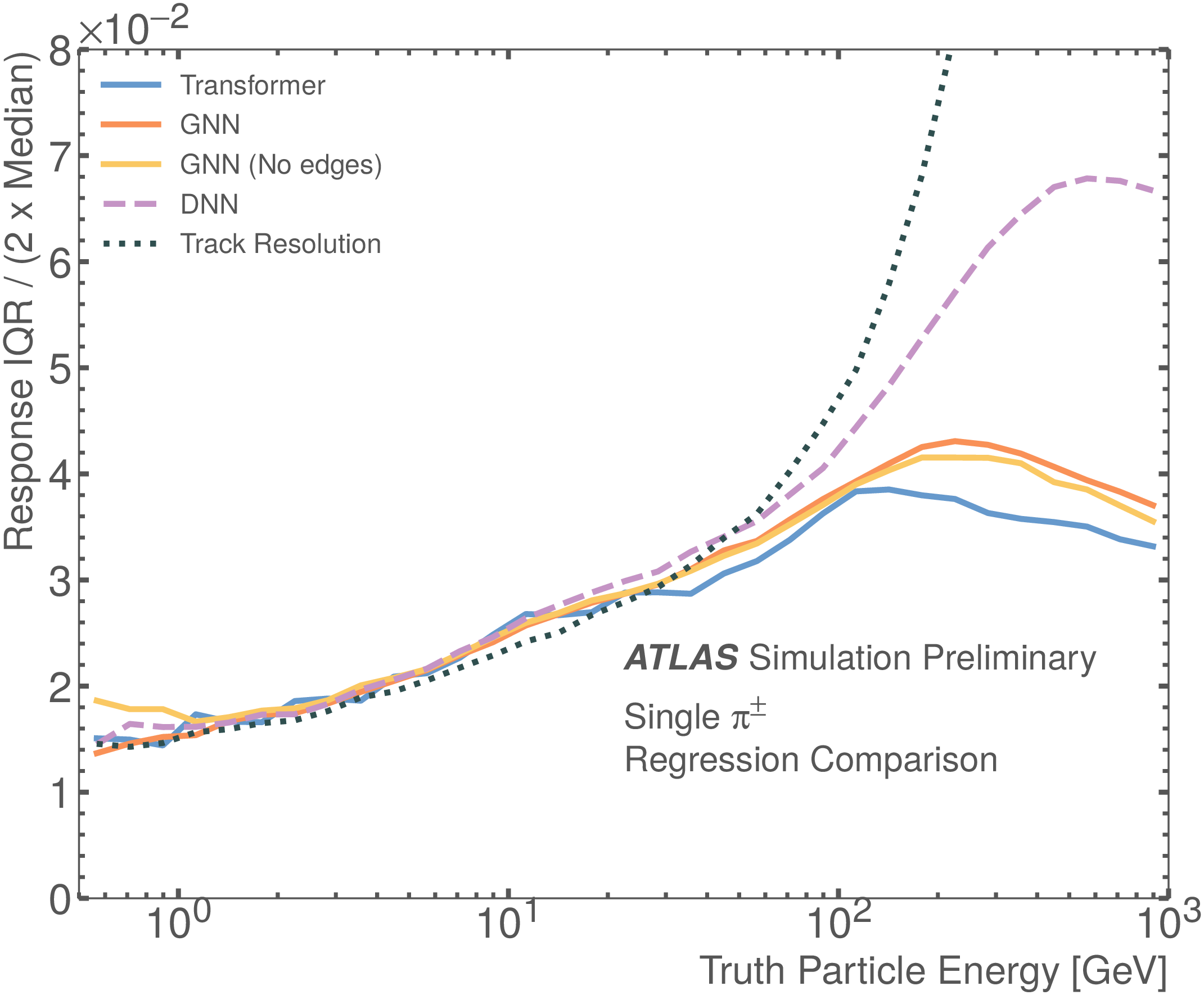}
	\label{fig:pion_res}}
\caption{		
	(a) Comparison of topo-cluster classification performance for three ML methods compared to the non-ML $\mathcal{P}^{\mathrm{EM}}_{\mathrm{clus}}$ baseline. (b) Resolution as a function of truth particle energy for several ML architectures alongside the track resolution. 
	Plots reproduced from Reference~\cite{ATL-PHYS-PUB-2022-040}.}
\label{fig:pion}
\end{figure}

The ATLAS detector has non-compensating calorimetry, so being able to distinguish charged and neutral pions allows the corresponding hadronic energy depositions to be restored to the correct scale.
Figure \ref{fig:pion_class} indicates the $\pi^0\ \mathrm{vs.}\ \pi^{\pm}$ classification performance of several ML models trained in Reference \cite{ATL-PHYS-PUB-2022-040} compared to a non-ML baseline classifier (labelled $\mathcal{P}^{\mathrm{EM}}_{\mathrm{clus}}$). 
All methods outperform the baseline and the Graph Neural Network (GNN) shows the best performance overall.
%the deposits from charged and neutral pions need to be restored to difference energy scales. 
%the most common component of hadronic jets, improving the classification accuracy of charged and neutral pions can therefore improve jet reconstruction performance. 

Additionally, several ML models are trained to calibrate pion energy. 
Given that pions are produced in abundance in nearly all hadronic showers, understanding and improving pion reconstruction is central to improving jet reconstruction. 
%This is considered a first step towards a larger goal of using ML in jet energy reconstruction. 
The models in Reference \cite{ATL-PHYS-PUB-2022-040} show significant improvement on current (non-ML) calibration baselines, particularly when combining tracking and calorimeter information. 
Figure \ref{fig:pion_res} shows energy resolution for several ML methods along with the track resolution. 
The resolution is quantified as one-half the interquantile range (IQR) divided by the median predicted energy, where the IQR represents the width of the response data (where response is the ratio of the predicted pion energy to particle-level pion energy) from 1$\sigma$ to $-1\sigma$ of the median. 
This captures a measure of the spread of energy predictions. 
The ML methods approximate the tracker energy resolution up to around $50\ \GeV$ before the calorimeter energy resolution dominates, indicating that ML is providing the best of both tracking and calorimetry performance. 

%Previous iterations of this study used image of calorimeter data as inputs, which excludes tracking data. 
%These results clearly indicate that the recent development to use point clouds, which include both topocluster cells and tracks, allow the ML models to use the best of both tracking and calorimeter information to produce a better calibration. 

\section{Boosted jet tagging}
\label{sec:tagging}

Large-radius jets from massive particles (such as $W$-bosons, $Z$-bosons and top quarks) can be distinguished from light quark/gluon-initiated jets using jet substructure information. 
ATLAS employs a variety of taggers for this purpose, and improving their classification accuracy enhances the performance of both searches and precision measurements. 
The latest taggers developed in ATLAS use jets reconstructed from Unified Flow Objects \cite{ATLAS:2020gwe} which combine topocluster and tracking information to provide improved pile-up resilience and jet mass resolution. 

\begin{figure}[b]
	\centering
	\subfloat[][]{
		\includegraphics[width=0.45\textwidth]{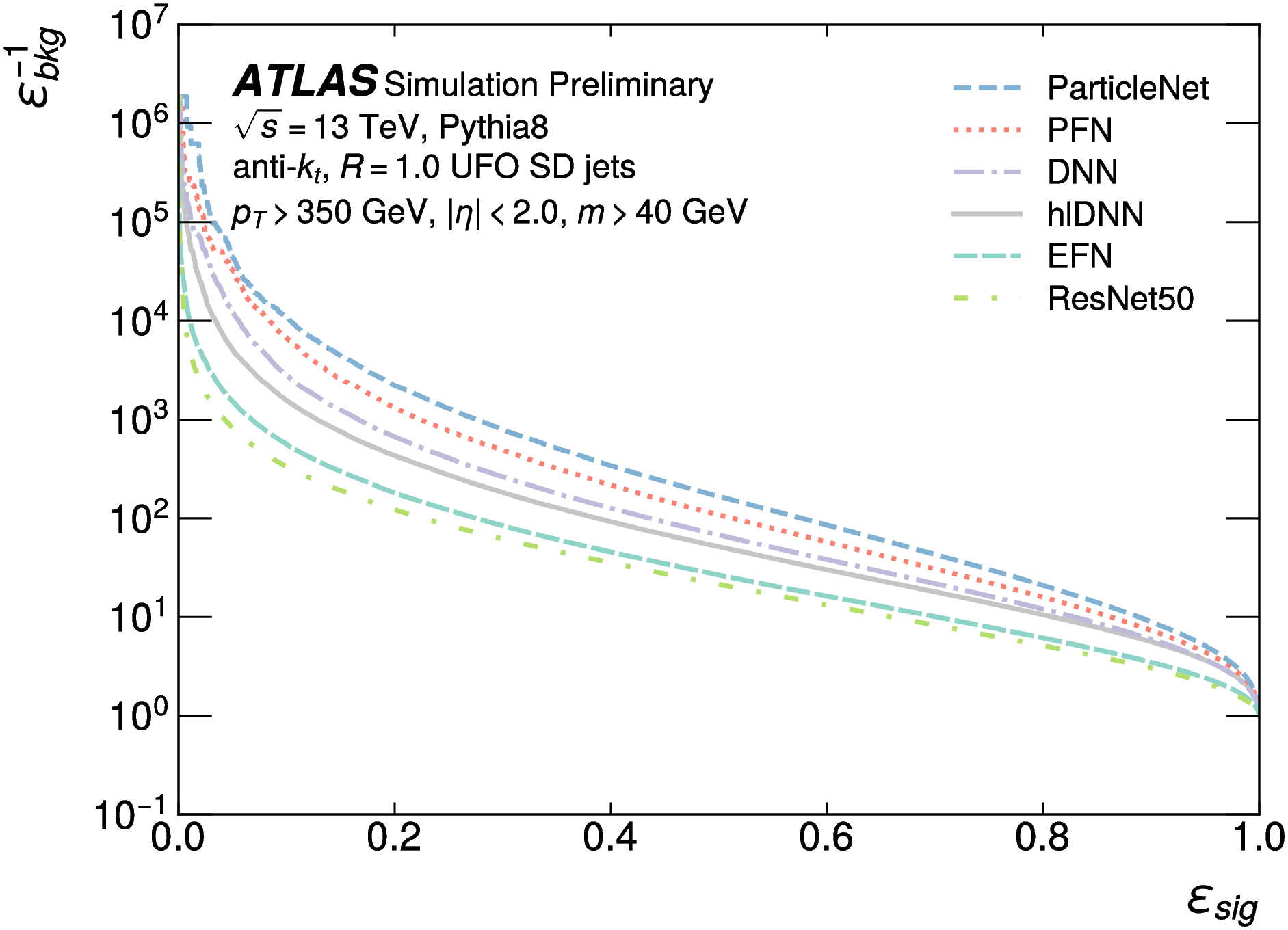}
		\label{fig:top_class}}
	\qquad
	\subfloat[][]{
		\includegraphics[width=0.45\textwidth]{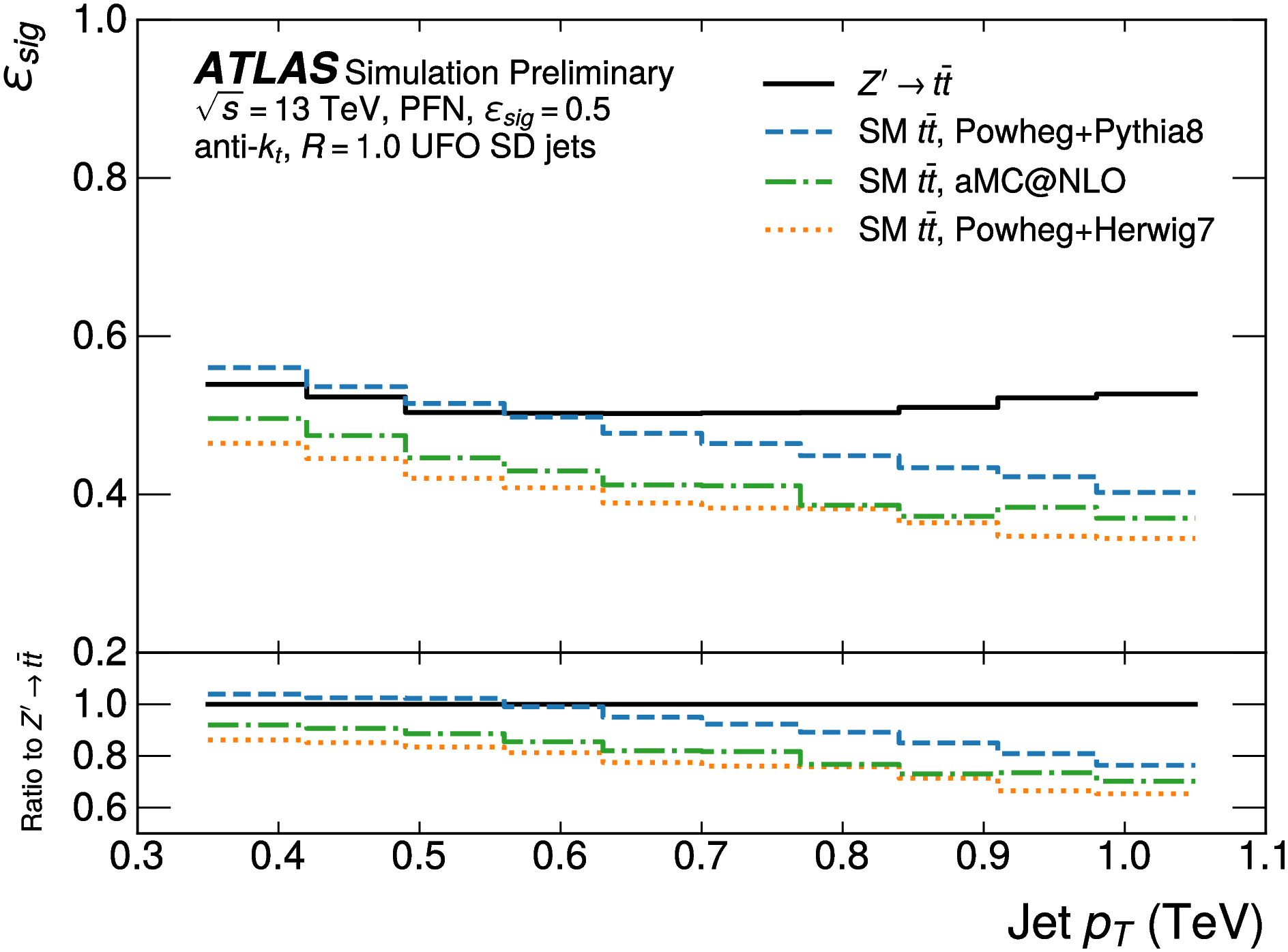}
		\label{fig:PNet_model}}
	\caption{
		(a) Classification efficiency for several top quark taggers. (b) The signal efficiency for ParticleNet for several \ttbar\ samples.
		Plots reproduced from Reference~\cite{ATL-PHYS-PUB-2022-039}.}
	\label{fig:top_taggers}
\end{figure}

In Reference \cite{ATL-PHYS-PUB-2022-039}, several top-quark taggers which were developed outside ATLAS using simplified Delphes simulated data-sets \cite{top_taggers_short} are evaluated using realistic GEANT4-simulated samples. 
Figure~\ref{fig:top_class} compares the efficiency of the taggers, which use constituent-based input information, to baseline deep neural networks trained on high-level (hDNN) and constituent-level (DNN) inputs.
ParticleNet and the Particle Flow Network (PFN) outperform the baselines, while the Energy Flow Network (EFN) and ResNet50 underperform relative to the previous Delphes-based studies in Reference \cite{top_taggers_short}. 
This highlights the need to develop taggers in a realistic context. 
As a result, the simulated data set used in this study has been made publicly available in Reference \cite{tagger_docs}.
%through the CERN open data portal \cite{cern_open_data}. Detailed documentation and training scripts are also available in Reference \cite{tagger_docs}.

In Figure~\ref{fig:PNet_model}, ParticleNet shows a dependence on the QCD modelling. This is also seen for PFN, but not for EFN due to its requirement of infra-red collinear safe inputs.
%This includes several architectures developed outside of ATLAS using simplified Delphes simulated data-sets \cite{10.21468/SciPostPhys.7.1.014}.
%Here their performance is evaluated using realistic GEANT4-simulated samples.
%ParticleNet \cite{Qu_2020}, Particle Flow Network (PFN) \cite{Komiske_2019}, Energy Flow Network (EFN) \cite{Komiske_2019} and ResNet50 \cite{he2015deep}. 

\begin{figure}[t]
	\centering
	\subfloat[][]{
		\includegraphics[width=0.45\textwidth]{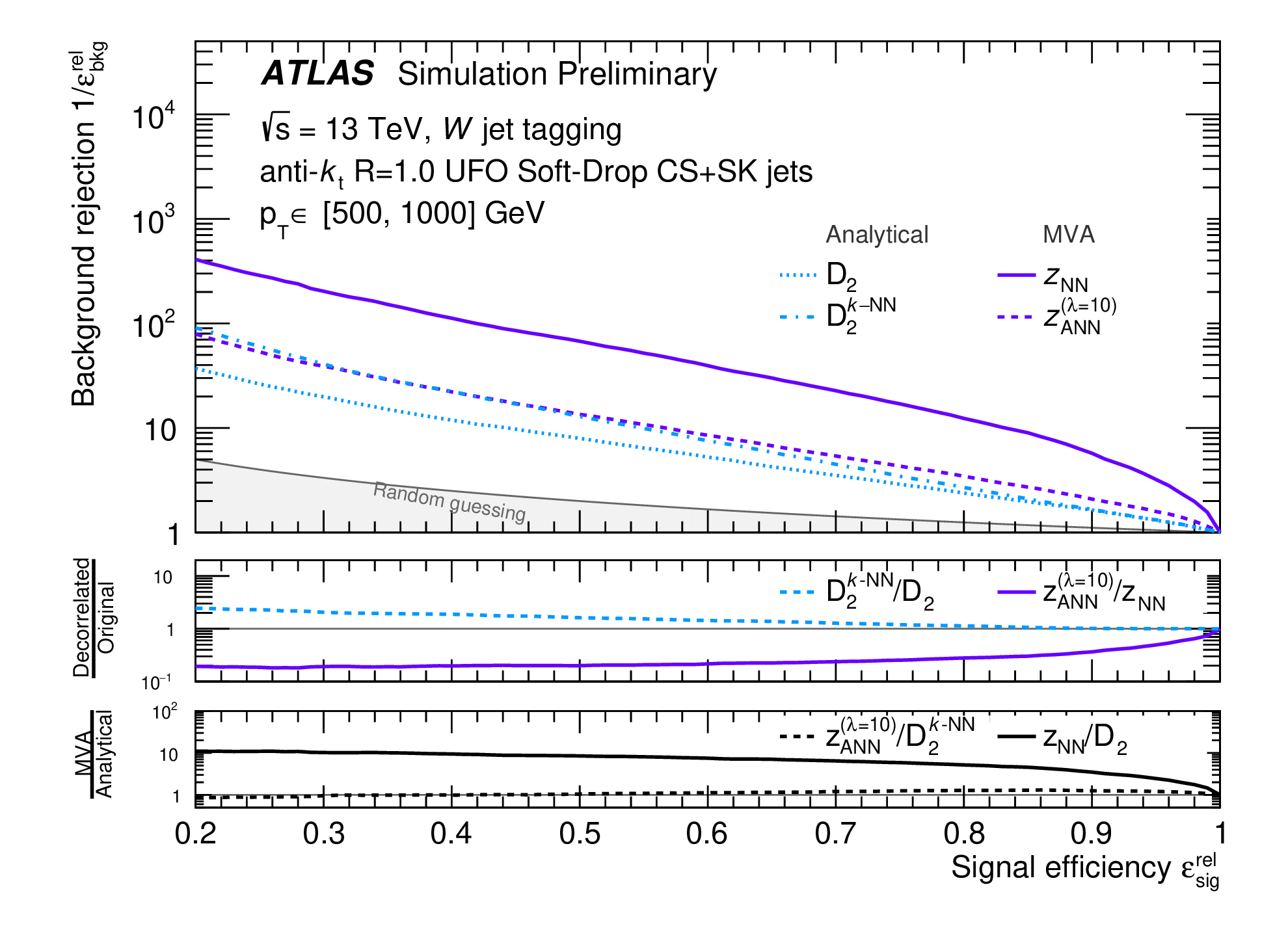}
		\label{fig:Wtag_eff}}
	\qquad
	\subfloat[][]{
		\includegraphics[width=0.45\textwidth]{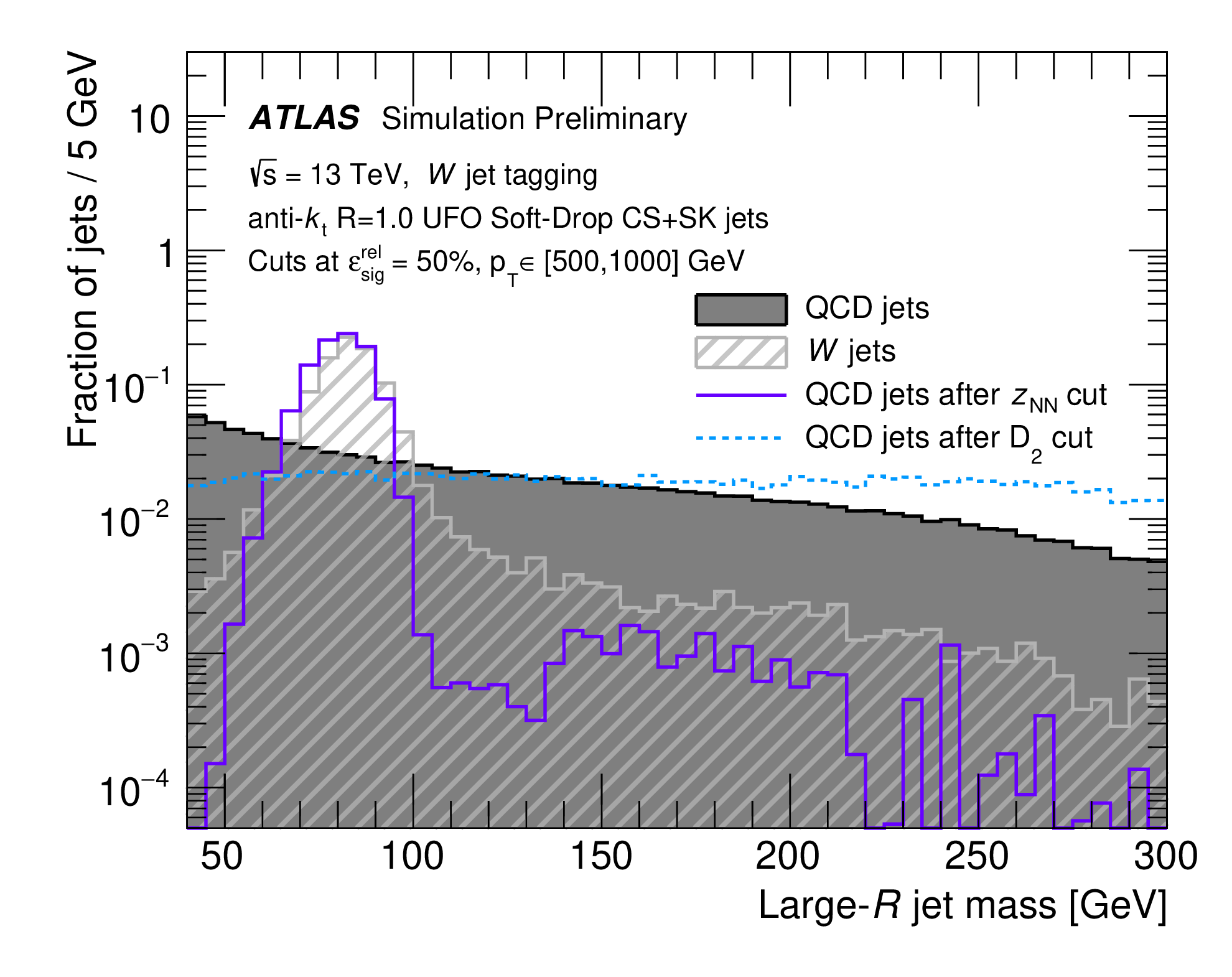}
		\label{fig:Wtag_dist}}	
	%\subfloat[][Mass decorrelation]{
	%	\includegraphics[width=0.45\textwidth]{W_tagging/Wtagger_massDecorr.png}
	%\label{fig:Wtag_decorr_dist}}
	\caption{
		(a) Classification efficiency for several $W$ taggers. (b) Jet mass distribution of $W$ jets signal and QCD background before and after tagger requirements. 
		Plots reproduced from Reference~\cite{ATL-PHYS-PUB-2021-029}.}
	\label{fig:Wtagger}
\end{figure}

References~\cite{ATL-PHYS-PUB-2021-029}~and~\cite{Wtag_update} include the latest developments in $W/Z$ tagging. 
A DNN shows improved performance compared to the current cut-based taggers, which can be seen by comparing $z_{\mathrm{NN}}$ with $D_2$ in Figure \ref{fig:Wtag_eff}. 
However, Figure \ref{fig:Wtag_dist} shows that the $z_{\mathrm{NN}}$ tagger sculpts the QCD background jet-mass distribution to match the $W$ jets signal topology, which would complicate background estimation strategies which use side-band regions. 
To address this, an adversarial NN (ANN) is trained to de-correlate the jet mass. 
This corresponds to a decrease in performance for the ANN, labelled $z_{\mathrm{ANN}}^{(\lambda=10)}$ in Figure \ref{fig:Wtag_eff}. 
The performance could partially be recovered with analysis-specific mass-window requirements.

\section{Conclusion}
\label{sec:Conclusion}

Hadronic object reconstruction at the LHC is ripe for ML applications. 
This contribution has presented some recent highlights in ATLAS, including regressing truth-level \ptmiss\ and pion energy and classifying pions and boosted jets. 
Development of all of these applications is ongoing and promises to enhance the performance of precision Standard Model measurements and beyond-the-Standard Model searches.

\bibliographystyle{JHEP}
{\footnotesize{\bibliography{ATLAS_ML-refs}}}

%\begin{thebibliography}{99}
%\bibitem{example_article}
%F. Baggins,
%\emph{Quantum effects of the One Ring},
%\href{https://doi.org/10.0000/3021006}
%{\emph{JHEP} \textbf{01} (3021) 006}
%[{\tt hep-th/2001033}].
%
%\bibitem{example_book}
%B. Baggins,
%\emph{There and back again},
%Imladris Editions, Rivendell 3018.
%
%\bibitem{example_contribution}
%F. Baggins,
%\emph{I will take the ring},
%in proceedings of \emph{Elrond Council},
%Imladris Editions, Rivendell 3021.
%
%\end{thebibliography}

\end{document}